\documentclass[12pt]{iopart}
\usepackage{iopams}
\usepackage{setstack}
\usepackage{amstext}
\usepackage{graphicx}
\usepackage{subfigure}
\usepackage{array}

\begin{document}

\title[]{Strategy for quantum algorithm design assisted by machine learning}
\author{Jeongho Bang$^{1,2}$, Junghee Ryu$^{3,2}$, Seokwon Yoo$^{2}$, Marcin Paw\l{}owski$^{3,4}$, and Jinhyoung Lee$^{2,1}$}
\address{$^1$ Center for Macroscopic Quantum Control \& Department of Physics and Astronomy, Seoul National University, Seoul, 151-747, Korea}
\address{$^2$ Department of Physics, Hanyang University, Seoul 133-791, Korea}
\address{$^3$ Institute of Theoretical Physics and Astrophysics, University of Gda\'{n}sk, 80-952 Gda\'{n}sk, Poland}
\address{$^4$ Department of Mathematics, University of Bristol, Bristol BS8 1TW, United Kingdom}
\ead{\mailto{jbang@snu.ac.kr}}
\ead{\mailto{hyoung@hanyang.ac.kr}}

\begin{abstract}
We propose a method for quantum algorithm design assisted by machine learning. The method uses a {\em quantum-classical hybrid} simulator, where a ``quantum student'' is being taught by a ``classical teacher.'' In other words, in our method, the learning system is supposed to evolve into a quantum algorithm for a given problem assisted by classical main-feedback system. Our method is applicable to design quantum oracle-based algorithm. As a case study, we chose an oracle decision problem, called a Deutsch-Jozsa problem. We showed by using Monte-Carlo simulations that our simulator {\em can} faithfully {\em learn} quantum algorithm to solve the problem for given oracle. Remarkably, learning time is proportional to the {\em square root} of the total number of parameters instead of the exponential dependance found in the classical machine learning based method.
\end{abstract}




\maketitle

\newcommand{\bra}[1]{\left<#1\right|}
\newcommand{\ket}[1]{\left|#1\right>}
\newcommand{\abs}[1]{\left|#1\right|}
\newcommand{\expt}[1]{\left<#1\right>}
\newcommand{\braket}[2]{\left<{#1}|{#2}\right>}
\newcommand{\ketbra}[2]{\left|{#1}\left>\right<{#2}\right|}
\newcommand{\commt}[2]{\left[{#1},{#2}\right]}

\newcommand{\identity}{1\!\!1}

\section{Introduction}\label{sec:1}

Quantum information science has seen explosive growth in recent years, as a more powerful generalization of classical information theory \cite{Nielsen99}. In particular, quantum computation has received momentum from the quantum algorithms that outperform their classical counterparts \cite{Deutsch85, Deutsch92, Shor97, Grover97}. Thus, the development of quantum algorithms is one of the most important areas of computer science. However, unfortunately, recent research on quantum algorithm design is rather stagnant, compared to other areas in quantum information, as new quantum algorithms have scarcely been discovered in the last few years \cite{Shor03}. We believe that this is due to the fact that we - the designers are used to {\em classical} logic. Thus we think that the quantum algorithm design should turn towards new methodology, different from that of the current approach. 

Machine learning is a well-developed branch of artificial intelligence and automatic control. Although ``learning'' is often thought of as a uniquely human trait, a machine being given feedback (taught) can improve its performance (learn) in a given task \cite{Uchiyama78, Langley95}. In the last decades, there has been a growing interest not only in the theoretical studies but also in a variety of applications of the machine learning. Recently, many quantum implementations of machine learning have been introduced to achieve better performance for quantum information processing \cite{Assion98,Sasaki02,Bisio10,Hentschel10,Bang12}. These works motivate us to look at machine learning as an alternative approach for quantum algorithm design.

Keeping our primary goal in mind, we ask whether a quantum algorithm can be found by the machine that also implements it. Based on this idea, we consider a machine which is able to learn quantum algorithms in a real experiment. Such a machine may discover solutions which are difficult for humans to find because of our classical way of thinking. Since we can always simulate a quantum machine on a classical computer (though not always efficiently) we can use such simulations to design quantum algorithms without the need for a programable quantum computer. This classical machine can thus be regarded as a simulator that learns a quantum algorithm, so-called {\em learning simulator}. The novelty of such a learning simulator is in its capabilities of ``learning'' and ``teaching.'' With regard to these abilities, we consider two internal systems: One is a learning system (``student'' say), and the other is a main feedback system (``teacher'' say). While the standard approach is to assume that both of student and teacher are quantum machines here we use a {\em quantum-classical hybrid} simulator such that the student is a quantum and the teacher a classical machine. Such a hybridization is easier and more economical to realize if any algorithms are able to be learned.

In this paper, we employ a learning simulator for quantum algorithm design. The main question of this work is: ``Can our learning simulator help in designing quantum algorithm?'' The answer to this question is affirmative, as it is shown, in Monte-Carlo simulations, that our learning simulator {\em can} faithfully {\em learn} appropriate elements of a quantum algorithm to solve an oracle decision problem, called Deutsch-Jozsa problem. The found algorithms are equivalent, but not exactly equal, to the original Deutsch-Jozsa algorithm. We also investigate the learning time, as it becomes important in application not only due to the large-scale problem often arises in machine learning but also for the fact that in its learning our simulator will exhibit the quantum speedup (if any) of an algorithm to be found, as described in later. We observe that the learning time is proportional to the {\em square root} of the total number of parameters, instead of the exponential tendency found in the classical machine learning. We expect that our learning simulator will reflect the quantum speedup of the found algorithm in its learning, possibly in synergy with the findings that the size of the parameter space can be significantly smaller for quantum algorithms than for their classical counterparts \cite{Manzano09}. We note that the presented method is aimed at a real experiment, in contrast to the techniques of \cite{Spector99, Behrman08}.


\section{Basic architecture of the learning simulator}\label{sec:2}

Before discussing the details of learning simulator, it is important to have an understanding of what machine learning is. A typical task of machine learning is to find a function $f(x)=t_x$ for the input $x$ and the target $t_x$ based on the observations in supervised learning or to find some hidden structure in unsupervised learning \cite{Uchiyama78, Langley95}. The main difference between supervised and unsupervised learning is that in latter case the target $t_x$ is unknown. Throughout this paper, we consider a supervised learning where the target $t_x$ is known.

\begin{figure}
\centering
\includegraphics[width=0.75\textwidth]{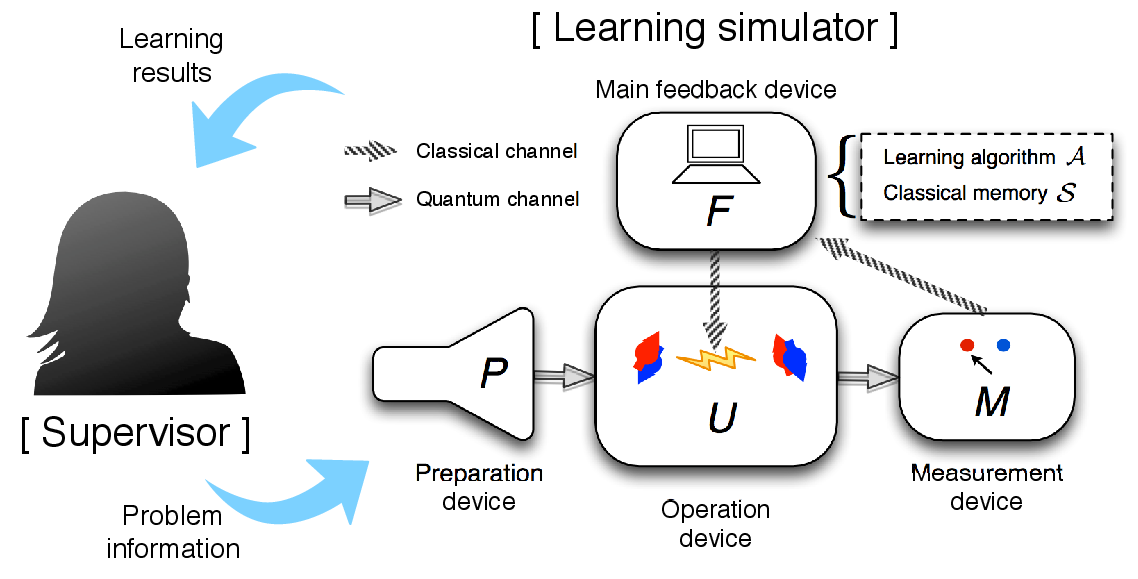}
\caption{Schematic picture of our method. A supervisor defines the problem to be solved and arranges the necessary prerequisites to learn quantum algorithm. All these information are communicated to the learning simulator at once. The simulator encodes these information on its own elements. The simulator consists of quantum elements, i.e. preparation $P$, operation $U$, and measurement $M$, assisted by classical main-feedback $F$. The classical channels ${\cal C}_{MF}$ and ${\cal C}_{FU}$ enable one-way communication from $M$ to $F$ and from $F$ to $U$.}
\label{fig:qlem}
\end{figure} 

We now briefly sketch our method (See also figure~\ref{fig:qlem}). To begin, a supervisor defines the problem to be solved, and arranges the necessary prerequisites (e.g., the input-target pairs ($x$, $t_{x}$), and a function $Q$ referred by a non-trivial device so-called oracle) for learning. These preliminary information are tossed to the learning simulator {\em at once}. The simulator encodes the communicated information on its own elements. We note here that one could consider two main issues in designing quantum algorithm. First is to construct a useful form of quantum oracle, and second is to find the other incorporating quantum operation(s) to maximize the quantum advantages, such as superposition engaging parallelism \cite{Cleve98} or entanglement \cite{Ekert98}. We here focus on the latter \footnote{Actually, in algorithm design \cite{Barnum04} or logic-mechanism programming \cite{Muggleton94}, the important point is usually that how we utilize a given oracle (or a corresponding operation to judge the positive or negative state) with other incorporated logics in order to achieve a speedup of the designed algorithm, rather than how we construct or optimize the oracle itself.}. Note, however, that it is necessary to define a specific oracle operation (See \ref{appendix_1}). This task is also performed, by the supervisor, at this preliminary stage. 


We then describe the basic elements of the learning simulator in figure~\ref{fig:qlem}. The simulator consists of two internal parts. One is the learning-system which is supposed to eventually perform a quantum algorithm, and the other is the feedback-system responsible for teaching the former. The learning-system consists of the standard {\em quantum} information-processing devices: Preparation $P$ to prepare a pure quantum state, operation $U$ to perform an unitary operation, and measurement $M$. Here, the chosen quantum oracle is involved in $U$. On the other hand, the feedback-system is {\em classical} as it is easier and less expensive to realize in practice. Furthermore, by employing the classical feedback, we can use a well-known (classical) learning algorithm whose performance has already been proved to be reliable. Recently, a scheme for machine learning involving a quantum feedback has been reported \cite{Gammelmark09}, but the usefulness of the quantumness has not been clearly elucidated, even though their results are meaningful in some applications. Moreover, it is unclear yet whether any classical feedback is applicable to the quantum algorithm design. Consequently, it is preferred to use the classical feedback in this work. In the sense, our simulator is a {\em quantum-classical hybrid}. The feedback-system is equipped with a main feedback device $F$ which involves the classical memory ${\cal S}$ and the learning algorithm ${\cal A}$. ${\cal S}$ records the control parameters of $U$ and measurement results of $M$. ${\cal A}$ corresponds to a series of rules for updating $U$.

We illustrate how our simulator performs the learning. Let us start with the set of $K$ input-target pairs communicated from the supervisor:
\begin{eqnarray}
T=\{(x_1, f(x_1)),(x_2, f(x_2)),\ldots,(x_K, f(x_K))\}, 
\end{eqnarray}
where $f$ is a function that transforms the inputs $x_i$ into their targets \footnote{Here, $x_i$ ($i = 1,2,\ldots,K$) can be encoded either on the state $\ket{\Psi_\text{in}}$ by $P$ or on the control parameters of $U$. In most cases, encoding on $U$ is appropriate and this is the case for our work, as shown later.}. The main task of the simulator is to find $f$. Firstly, an initial state $\ket{\Psi_\text{in}}$ is prepared in $P$ and transformed to $\ket{\Psi_\text{out}}$ by $U$. Then $M$ performs measurement on $\ket{\Psi_\text{out}}$ with a chosen measurement basis. The measurement result is delivered to $F$ through ${\cal C}_{MF}$. Note here that the information about the initial state $\ket{\Psi_\text{in}}$ and the measurement basis encoded in $P$ and $M$ are also determined by the supervisor before the learning. Finally, $F$ updates $U$ based on ${\cal A}$. Basically, the learning is just the repetition of these three steps. When the learning is completed, we obtain $P$-$U$-$M$ device to implement $f$ by simply removing $F$. The supervisor, then, investigate if the found $P$-$U$-$M$ provides any speedup reducing the overall oracle references, or saves any computational resources to implement the algorithm \cite{Cleve99}. In particular, the supervisor would standardize the identified operations $U$ as an algorithm. Here, we clarify that the input information in $T$ and the measurement results are classical.  Nevertheless, the simulator is supposed to exploit quantum effects in learning, because the operations before measurement are all quantum. This assumption is supported by recent theoretical studies that show the improvement of the learning efficiency by using quantum superposition \cite{Manzano09, Yoo13}.


\section{Construction of the learning simulator}\label{sec:3}

\begin{figure}
\centering	
\includegraphics[width=0.7\textwidth]{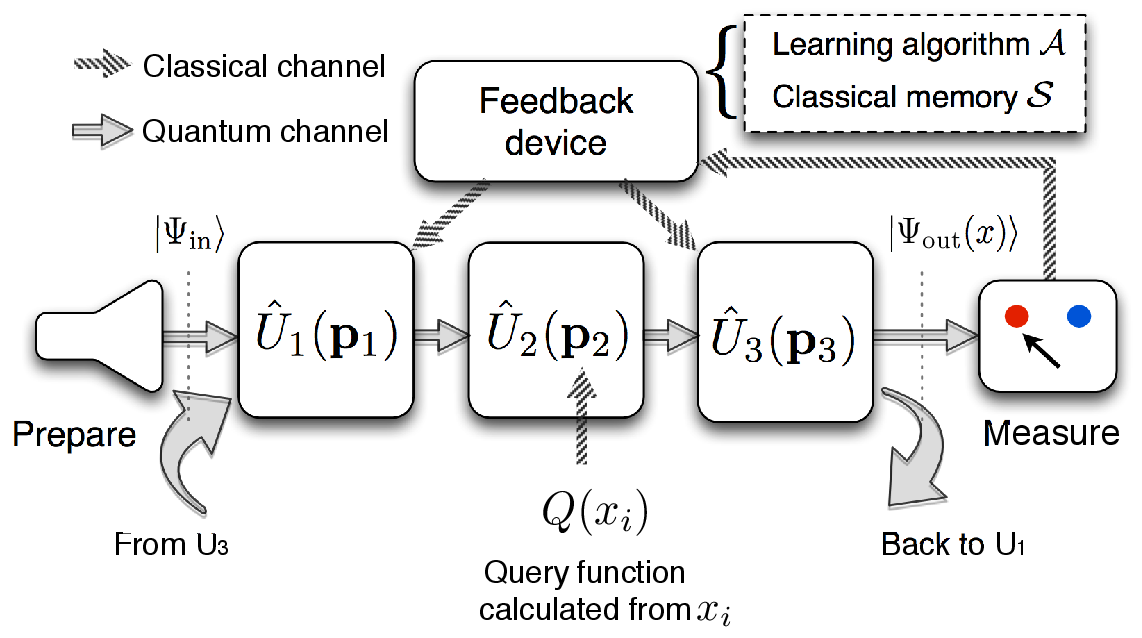}
\caption{Architecture of our simulator to learn a quantum algorithm, where the unitary operation $U$ consists of three sub-operations (See the text).}
\label{fig:qlemda}
\end{figure}

The general design of the learning simulator depicted in figure~\ref{fig:qlem} works fine for problems, such as number factorization. However, in the problems requiring a large number of oracle references, the input is the oracle itself and, by definition, it is a (unitary) transformation rather than a string of bits. To allow for the input in the form of an unitary matrix we need to refine our simulator a little (but let us stress that this does not mean that our method is not general). The refined version depicted in figure~\ref{fig:qlemda} allows the simulator to learn an algorithm of iterative type. The difference in the learning simulators stems directly from the formulation of the problems.

The most important aspect in the refined learning simulator is the decomposition of $U$. In order to deal with both classical and quantum information, we divide $U$ into three sub-devices, such that
\begin{eqnarray}
\label{eq:unitop}
\hat{U}_\text{tot} = \hat{U}_3(\mathbf{p}_3) \hat{U}_2(\mathbf{p}_2) \hat{U}_1(\mathbf{p}_1),
\end{eqnarray}
where $\hat{U}_{tot}$ is total unitary operator, and $\hat{U}_j$ ($j=1,2,3$) denotes the unitary operator of $j$th sub-device. Here, $\hat{U}_1$ and $\hat{U}_3$ are $n$-qubit controllable unitary operators, whereas $\hat{U}_2$ is the oracle to encode the input $x_i$. By `controllable' we here, and throughout the paper, means that they can be changed by the feedback.

The unitary operators are {\em generally} parametrized as 
\begin{eqnarray}
\hat{U}(\mathbf{p}) = \exp{\left(-i \mathbf{p}^T \mathbf{G}\right)}, 
\label{eq:u_p}
\end{eqnarray}
where $\mathbf{p}=(p_1,p_2,\ldots,p_{d^2-1})^T$ is a real vector in $(d^2-1)$-dimensional Bloch space for $d=2^n$, and $\mathbf{G}=(\hat{g}_1,\hat{g}_2,\ldots\hat{g}_{d^2-1})^T$ is a vector whose components are SU($d$) group generators  \cite{Hioe81,Son04}. The components $p_j \in [-\pi,\pi]$ of $\mathbf{p}$ can directly be matched to control parameters in some experimental schemes, e.g., beam-splitter and phase-shifter alignments in linear optical system \cite{Reck94} or radio-frequency (rf) pulse sequences in nuclear magnetic resonance (NMR) system \cite{Lee00}. In that sense, we call $\mathbf{p}$ a control-parameter vector. Here, $\mathbf{p}_2$ is determined by $Q(x_i) \mapsto \mathbf{p}_2(x_i)$, as described above. In such setting, we expect that our simulator learns an optimal set of $\{\mathbf{p}_{1}, \mathbf{p}_{3}\}$, so that $\hat{U}_1$ and $\hat{U}_3$ come to solve a given problem.

Our simulator is actually well-suited to learn even iterative algorithms, such as Grover's \cite{Grover97}. We envision using our simulator as follows: In the first stage, apply $\hat{U}_1$ to an input-state, then $\hat{U}_2$ which is a non-trivial operation, say oracle, and finally $\hat{U}_3$ to generate an output state. The feedback-system updates $\hat{U}_1$ and $\hat{U}_3$. Then, after a certain number of iterations which do not lead to any improvements, our simulator goes to the second stage, where the output state is fed back to be the input state to apply $\hat{U}_1$-$\hat{U}_2$-$\hat{U}_3$ again. Therefore, in the second stage, the oracle is referenced twice. If it fails again, it will try to loop three times at the third stage. By some number of stages, there will be enough oracle references to solve the problem. In such way, our simulator can learn even a quantum algorithm of iterative type \footnote{The procedure is not the most general one. For full generality one would also need to add some quantum memory but, to our knowledge, no existing quantum algorithm actually uses it yet.}, without adopting any additional sub-devices and altering the structure in a real experiment. Thus, the scalability for the size of the search space is only concerned with the number of control parameters in $\hat{U}_{1}$ and $\hat{U}_3$, given by $D = 2 (d^2-1)$, where $d=2^n$.


Here, we highlight another subsidiary question: How long does it take for our simulator to learn a (almost) deterministic quantum algorithm? Investigating this issue will be increasingly important, especially in application of our simulator to very large-scale (i.e. $D \gg 1$) problem. Thus one may doubt that our simulator runs extremely slow in a large size of problem, on the one hand. On the other hand, however, it is also likely that in its learning our simulator enjoys the quantum speedup, if any, of an algorithm to be found. To see this, consider two cases, a classical and a quantum algorithm which our simulator tries to find, assuming that they are of different complexities in terms of the number of oracle queries. For instance, the quantum queries a polynomial number of oracles, whereas the classical does the exponential with respect to the problem size. Regardless of its realization methods, a learning simulator can reduce the number of stages not less than the number of oracle queries in a given algorithm to be found. This is reflected by learning time. In other words, our simulator may show the learning speedup, exploring much less stages in the learning of quantum algorithm, as far as the algorithm to be found exhibits quantum speedup. These controversial arguments demand us to investigate the learning time as well as the effectiveness of our simulator.


\section{Application to Deutsch-Jozsa problem}\label{sec:4}

As a case study, consider an $n$-bit oracle decision problem, called Deutsch-Jozsa (DJ) problem. The problem is to decide if some binary function $x_i$:$\{0,1\}^n \rightarrow \{0,1\}$ is constant ($x_i$ generates the same value $0$ or $1$ on every input) or balanced ($x_i$ generates  $0$ on exactly half of the inputs, and $1$ on the rest of the inputs) \cite{Deutsch85,Deutsch92}. On a classical Turing machine  $2^{n-1}+1$ queries are required to solve this problem. If we use a probabilistic random classical algorithm, we can determine the function $x_i$ with a small error, less than $2^{-q}$, by $q$ queries \cite{Arvind03,Adcock09}.

On the other hand, DJ quantum algorithm solves the problem by only single query \cite{Collins98,Adcock09}. The DJ quantum algorithm runs as follows: First, apply $\hat{H}^{\otimes n}$ on the input state $\ket{\Psi_\text{in}}=\ket{00 \cdots 0}$, then $\hat{U}_x$ to evaluate the input function, and finally $\hat{H}^{\otimes n}$ again to produce an output state $\ket{\Psi_\text{out}}$. Here, $\hat{H}$ is Hadamard gate which transforms the qubit states $\ket{0}$ and $\ket{1}$ into equal superposition states $\hat{H}\ket{0}=\left(\ket{0}+\ket{1}\right)/\sqrt{2}$ and $\hat{H}\ket{1}=\left(\ket{0}-\ket{1}\right)/\sqrt{2}$ respectively. $\hat{U}_x$ is the function-evaluation gate that calculates a given function $x_i$. It is defined by its action, 
\begin{eqnarray}
\hat{U}_x\ket{k_1 k_2 \cdots k_n} = e^{i\pi x_i(k_1 k_2 \cdots k_n)}\ket{k_1 k_2 \cdots k_n}, 
\label{eq:ux}
\end{eqnarray}
where $k_1 k_2 \cdots k_n \in \{0,1\}^n$ is the binary sequence of the computational basis. Then, the output state is given as
\begin{eqnarray}
\label{eq:out_st}
\ket{\Psi_\text{out}(x_i)}=
\left\{
\begin{array}{ll}
\pm \ket{00 \cdots 0}, & \text{if} ~~x_i \in C \\
\pm \ket{z_1 z_2 \cdots z_{n}}, & \text{if} ~~x_i \in B
\end{array}
\right.
\end{eqnarray}
where $C$ and $B$ are the sets of constant and balanced functions, respectively, and the binary components $z_j \in \{0, 1\}$ ($j=1,2,\ldots,n$) depend on the $d \choose d/2$ balanced functions (excepting that $z_j = 0$ for all $j$). In the last step, von-Neumann measurement is performed on the output state. The corresponding measurement operator is given by $\hat{M} = \ketbra{00 \cdots 0}{00 \cdots 0}$. The other projectors constituting the observable are irrelevant because we are interested only in the probabilities associated with the first case
\begin{eqnarray}
P_C = \bra{\Psi_\text{out}(x_i)}\hat{M}\ket{\Psi_\text{out}(x_i)}=1, ~~\text{if} ~~x_i \in C,
\label{eq:pc}
\end{eqnarray}
and the second case
\begin{eqnarray}
\label{eq:pb}
P_B = \bra{\Psi_\text{out}(x_i)}\hat{M}\ket{\Psi_\text{out}(x_i)}=0, ~~\text{if} ~~x_i \in B.
\end{eqnarray}
Therefore it is promised that the function $x_i$ is either constant or balanced by only single oracle query.

We are now ready to apply our method to the DJ problem. To begin, supervisor prepares the set of input-target pairs, $T=\{ (x_i, f(x_i)) | f(x_i)=\text{`c'} ~\text{if}~ x_i \in C ~\text{and}~ f(x_i)=\text{`b'} ~\text{if}~ x_i \in B \}$. The learning simulator is to find the ``functional'' $f$ now as adjusting $\hat{U}_1$ and $\hat{U}_3$. The input functions $x_i$ are encoded in $\mathbf{p}_2(x_i)$ of $\hat{U}_2$. Here, we chose the same form of the oracle as equation~(\ref{eq:ux}), i.e. type ($ii$). Then $P$ prepares an {\em arbitrary} initial state $\ket{\Psi_\text{in}}$ and $M$ performs the measurement on each qubit. Here we introduce a function to apply a measurement result to one of the targets (in our case, `c' or `b'). We call this interpretation function. Note that the interpretation function is also to be learned, because, in general, any {\em a priori} knowledge of the quantum algorithm to be found is completely unknown. For a sake of convenience, we consider a Boolean function that transforms the measurement result $z_1 z_2 \cdots z_n$ to $0$ (equivalently, `c') only if $z_j=0$ for all $j=1,2,\ldots,n$, and otherwise $1$ (equivalently, `b'). One may generalize the interpretation function to a function $\{0,1\}^n \rightarrow \{0,1\}^m$, if interested in any other problems that contain many targets less than $2^m$ \cite{Toffoli80}.


\section{Learning algorithm of differential evolution}\label{sec:5}

One of the most important parts in our method is choosing a learning algorithm ${\cal A}$. Efficiency and accuracy of machine learning are heavily influenced in general by the algorithm chosen. We employ so-called ``differential evolution'', as it is known as one of the most efficient optimization methods \cite{Storn97}. We implement the differential evolution as follows. To begin, we prepare $N_\text{pop}$ sets of the control parameter vectors: $\{\mathbf{p}_{1,i}, \mathbf{p}_{3,i}\}$ ($i=1,2,\cdots,N_\text{pop}$). Thus we have $2N_\text{pop}$ parameter vectors in total. They are chosen initially at random and recorded on ${\cal S}$ in $F$. [$\mathbf{L.1}$] Then, $2N_\text{pop}$ mutant vectors $\boldsymbol\nu_{k,i}$ are generated for $\hat{U}_k$ ($k=1,3$), according to 
$$
\boldsymbol\nu_{k,i} = \mathbf{p}_{k,a} + W \left(\mathbf{p}_{k,b} - \mathbf{p}_{k,c}\right),
$$
where $\mathbf{p}_{k,a}$, $\mathbf{p}_{k,b}$, and $\mathbf{p}_{k,c}$ are randomly chosen for $a,b,c \in \{1,2,\cdots,N_\text{pop}\}$. These three vectors are chosen to be different from each other, for that $N_\text{pop} \ge 3$ is necessary. The free parameter $W$, called a differential weight, is a real and constant number. [$\mathbf{L.2}$] After that, all $2N_\text{pop}$ parameter vectors 
$$
\mathbf{p}_{k,i}=(p_{k,1}, p_{k,2}, \cdots, p_{k,d^2-1})_i^T
$$ 
are reformed to trial vectors 
$$
\boldsymbol\tau_{k,i}=(\tau_{k,1}, \tau_{k,2}, \cdots, \tau_{k,d^2-1})_i^T
$$
by the rule: For each $j$,
\begin{eqnarray}
\label{eq:crossover}
\left\{
\begin{array}{ll}
\tau_{k,j} \leftarrow p_{k,j} & ~~\text{if}~R_j > C_r,\\
\tau_{k,j} \leftarrow \nu_{k,j} & ~~\text{otherwise}, \\
\end{array}
\right.
\end{eqnarray}
where $R_j \in [0, 1]$ is a randomly generated number and the crossover rate $C_r$ is another free parameter in between $0$ and $1$. [$\mathbf{L.3}$] Finally, $\{\boldsymbol\tau_{1,i}, \boldsymbol\tau_{3,i}\}$ are taken for the next iteration if $\hat{U}_{1}(\boldsymbol\tau_{1,i})$ and $\hat{U}_{3}(\boldsymbol\tau_{3,i})$ yield a larger fitness value than that from $\hat{U}_{1}(\mathbf{p}_{1,i})$ and $\hat{U}_{3}(\mathbf{p}_{3,i})$; if not, $\{\mathbf{p}_{1,i}, \mathbf{p}_{3,i}\}$ are retained. Here the fitness $\xi_i$ is defined by
\begin{eqnarray}
\label{eq:fitness}
\xi_i = \frac{P_{C,i} + \left(1-P_{B,i}\right)}{2},
\end{eqnarray}
where $P_{C,i}$ and $P_{B,i}$ are measurement probabilities for $i$-th set, given by equations~($\ref{eq:pc}$) and ($\ref{eq:pb}$). While evaluating the $N_\text{pop}$ fitness values, $F$ records on ${\cal S}$ the best $\xi_\text{best}$ and its corresponding parameter vector set $\{\mathbf{p}_{1,\text{best}}, \mathbf{p}_{3,\text{best}}\}$. The above steps [$\mathbf{L.1}$]-[$\mathbf{L.3}$] are repeated until $\xi_\text{best}$ reaches close to $1$. In an ideal case, the simulator finds $\{\mathbf{p}_{1,\text{best}}, \mathbf{p}_{3,\text{best}}\}$ that yields $\xi_\text{best}=1$ with $P_C=1$ and $P_B=0$. The found parameters lead to an algorithm equivalent to the original DJ.


\section{Numerical analysis}\label{sec:6}

\begin{figure}[t]
\centering
\includegraphics[angle=270,width=0.35\textwidth]{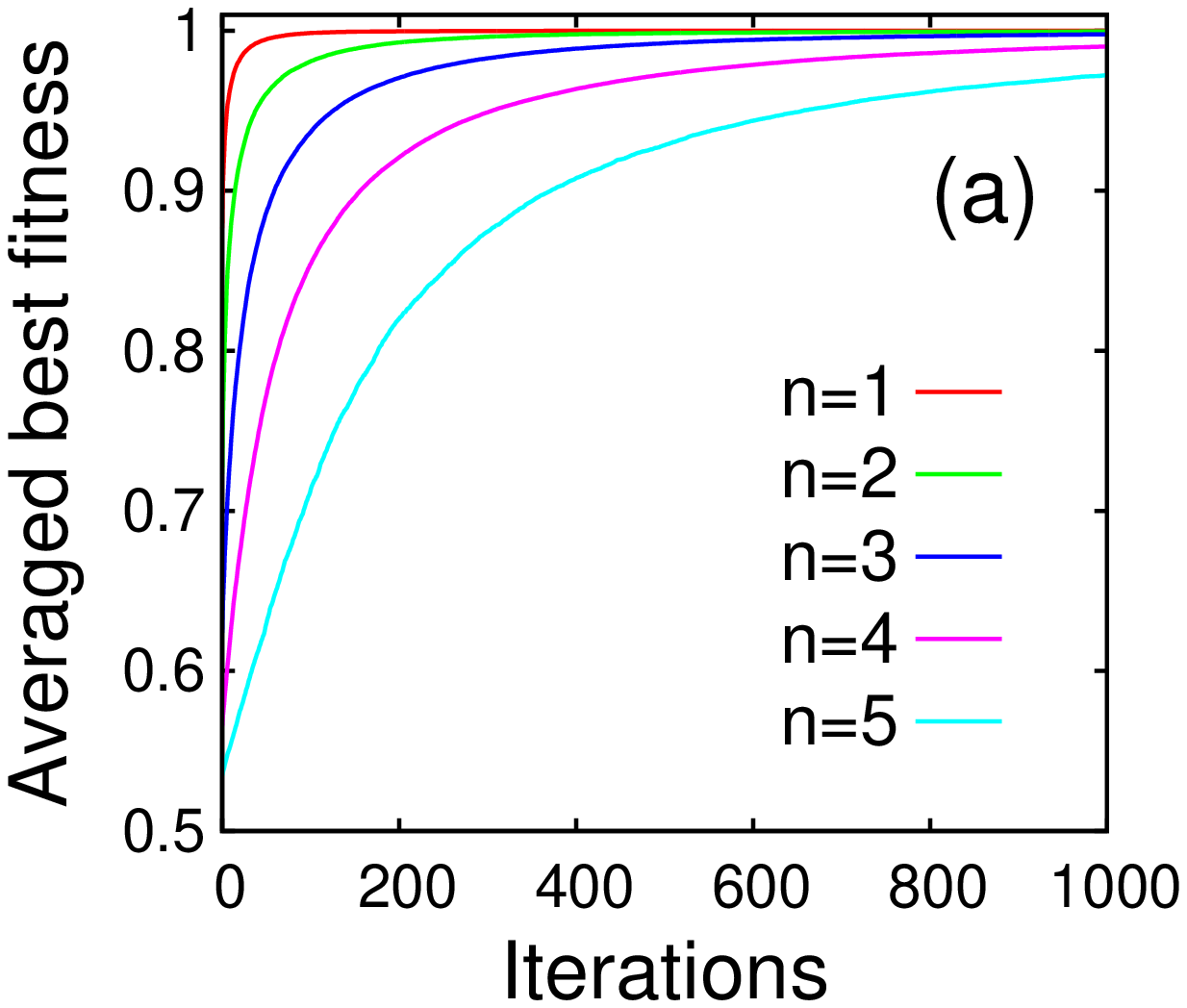} 
\includegraphics[angle=270,width=0.35\textwidth]{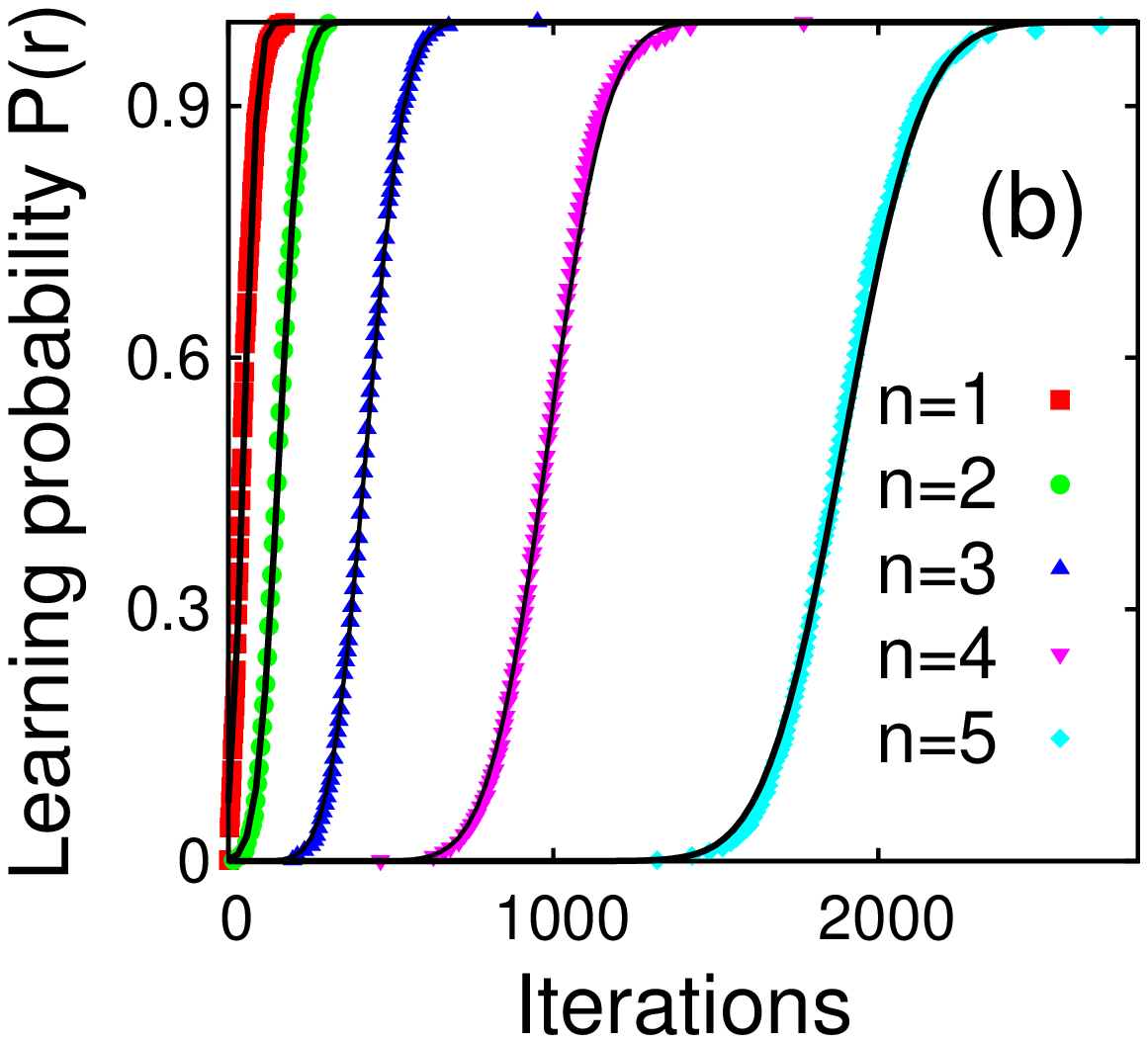}
\\
\includegraphics[angle=270,width=0.35\textwidth]{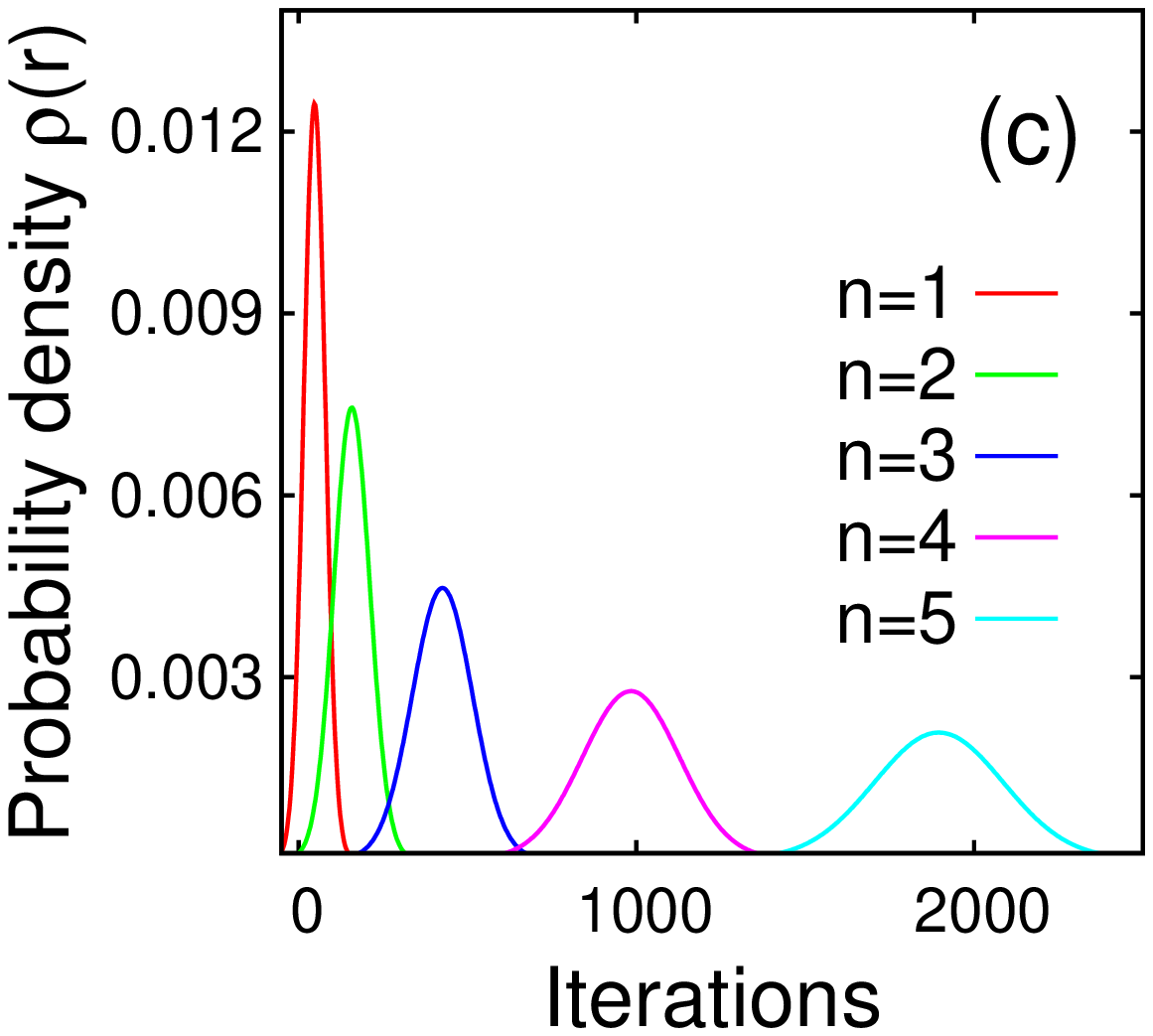} 
\includegraphics[angle=270,width=0.35\textwidth]{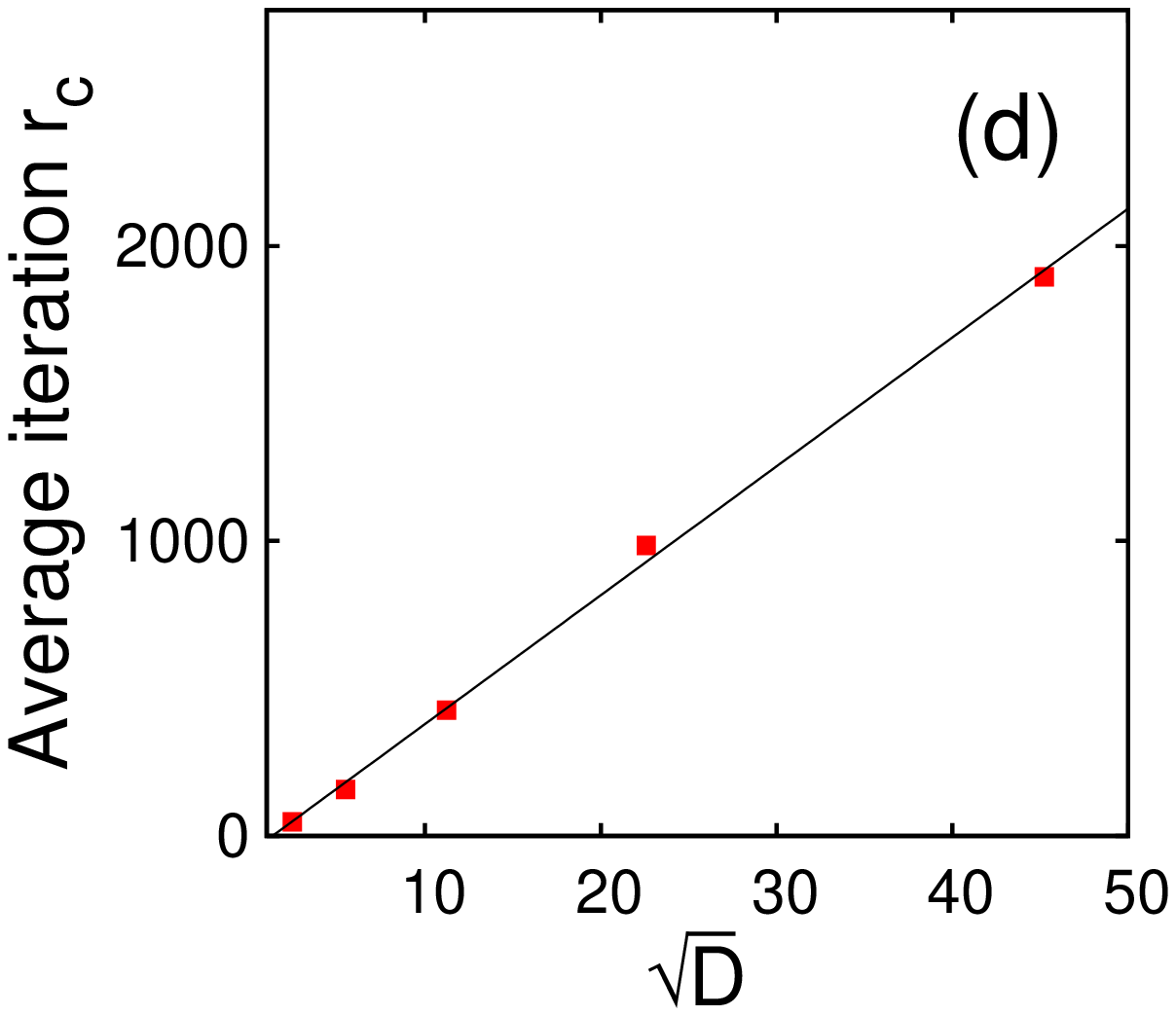}
\caption{(a) Averaged best fitness $\overline{\xi}_\text{best}$ versus iteration $r$. Each data is averaged over $1000$ simulations. It is observed that $\overline{\xi}_\text{best}$ approaches unity as iterating. (b) Learning probability $P(r)$ in the halting condition $\xi_\text{best} \ge 0.99$, sampling $1000$ trials. $P(r)$ is well fitted to an integrated Gaussian (black solid line), $G(r)=\int_{-\infty}^{r} dr' \rho(r')$. (c) Probability density $\rho(r)$ resulting from $P(r)$ for each $n$. (d) Graph of $r_c$ versus $\sqrt{D}$, where $D$ is the total number of the control parameters, and $r_c$ is the average number of iterations to complete the learning. The data are well fitted linearly to $r_c=A\sqrt{D}+B$ with $A \simeq 43$ and $B \simeq -57$.}
\label{grp:sim_dj}
\end{figure}

The simulations are done for $n$-bit DJ problem with increasing $n$ from $1$ to $5$. In the simulations, we take $N_\text{pop}=10$ for all $n$ \footnote{For a large size of classical learning-system, huge number $N_\text{pop}$ of candidate solutions are usually needed. For example, it is appropriate to chose $N_\text{pop} \simeq 5D \sim 10D$ (See the reference \cite{Storn97}).}. The results are given in figure~\ref{grp:sim_dj}(a), where we present the averaged best fitness $\overline{\xi}_\text{best}$, sampling $1000$ trials. It is clear to observe that $\overline{\xi}_\text{best}$ approaches to $1$ as iteration proceeds. The required stage is just one for all $n$. This implies that our simulator {\em can} faithfully {\em learn} a single-query quantum algorithm for DJ problem, showing $\xi \simeq 1$. It is also notable that the found algorithms are equivalent to, but not exactly equal to the original DJ algorithm: The found $\hat{U}_1$ and $\hat{U}_3$ are always different, but constitute an algorithm solving DJ problem (See \ref{appendix_2}).

Then we present a learning probability $P(r)$, defined by the probability that the learning is completed before or at $r$-th iteration \cite{Bang08}. Here we assume a halting condition $\xi_\text{best} \ge 0.99$ to find a {\em nearly deterministic} algorithm. In figure~\ref{grp:sim_dj}(b), we present $P(r)$ for all $n$, each of which is averaged by $1000$ simulations. We find that $P(r)$ is well fitted to an integrated Gaussian 
\begin{eqnarray}
G(r) = \int_{-\infty}^{r} dr' \rho(r'),
\end{eqnarray}
where probability density $\rho(r)$ is a Gaussian function $\frac{1}{\sqrt{2\pi}{\Delta r}} e^{- \frac{(r-r_c)^2}{2 {\Delta r}^2}}$. Here, $r_c$ is the average iteration number and ${\Delta r}$ is the standard deviation over the simulations, which characterize how many iterations are sufficient for a statistical accuracy of $\overline{\xi}_\text{best} \ge 0.99$. Note that we have {\em finite} values of $r_c$ and ${\Delta r}$ for all $n$. The probability density $\rho(r)$ is drawn in figure~\ref{grp:sim_dj}(c), resulting from $P(r)$.

We also investigate the learning time. As we already pointed out, learning time becomes an intriguing issue which may be related not only to the applicability of our algorithm to large-scale problem but also to the learning speedup. Regarding $r_c$ as a learning time, we present the graph of $r_c$ versus $\sqrt{D}$ in figure~\ref{grp:sim_dj}(d). Remarkably, the data are well fitted {\em linearly} to $r_c = A\sqrt{D} + B$ with $A \simeq 43$ and $B \simeq -57$. This means that the learning time is proportional to the {\em square root} of the size of the parameter space \footnote{It is worth noting that there is an alternative method, called semidefinite programming, which may be used for the purpose of finding a quantum algorithm. In reference \cite{Barnum04}, the authors have considered the problem of finding optimal unitaries given a fixed number of queries. Their algorithm could solve the problem in polynomial time (i.e. polynomial in the dimension $d$).}. This behavior is contrary to a typical tendency of being exponential in the classical machine learning (See, for example, \cite{Bergh04,Chu11} and their references). 	


\section{Summary and remarks}\label{sec:7}

We have presented a method for quantum algorithm design based on machine learning. The simulator we have used is a {\em quantum-classical hybrid}, where the quantum student is being taught by a classical teacher. We discussed that such a hybridization is beneficial in terms of the usefulness and the implementation cost. Our simulator was applicable to design an oracle-based quantum algorithm. As a case study, we demonstrated that our simulator can faithfully learn a single-query quantum algorithm that solves DJ problem even though it does not have to. The found algorithms are equivalent, but not exactly equal, to the original DJ algorithm with the fitness $\simeq 1$. 

We also investigated the learning time, as it would become increasingly important in application not only due to the large-scale problem often arises in machine learning but also for the fact that in its learning our simulator potentially exhibits the quantum speedup, if any, of an algorithm to be found. In the investigation, we observed that the learning time is proportional to the {\em square root} of the size of the parameter space instead of the exponential dependance in the classical machine learning. This result is very suggestive. We expect that our simulator will reflect the quantum speedup of the found algorithm in its learning, possibly in synergy with the findings from the reference~\cite{Manzano09} that for quantum algorithms the size of the parameter space can be significantly smaller than for their classical counterparts: Not only their learning time scales more favorably with the size of the space but also this size is smaller to begin with. 

We hope that the proposed method will help in designing quantum algorithms, and provide an insight of learning speedup establishing the link between the learning time and the quantum speedup of the found algorithms. Nevertheless, it is an open question whether to observe more improved behaviors in quantum algorithm design when employing a quantum feedback, compared with the classical feedback.


\section*{Acknowledgments}

J.B. would like to thank M. \.{Z}ukowski, H. J. Briegel, and B. C. Sanders for discussions and comments. We acknowledge the financial support of the National Research Foundation of Korea (NRF) grant funded by the Korea government (MEST) (No. 2010-0018295 and No. 2010-0015059). J. R. and M. P. are supported by the Foundation for Polish Science TEAM project cofinanced by the EU European Regional Development Fund. J.R. is also supported by NCBiR-CHIST-ERA Project QUASAR. M. P. is also supported by UK EP-SRC and ERC grant QOLAPS. 

\appendix

\section{Quantum oracle operation}\label{appendix_1}

As described in the main text, one could consider two different issues in designing a certain type quantum-algorithm. First is to determine a specific form of quantum oracle operation, and second is to find the other incorporating operations to maximize the quantum advantages. Although we focused on the latter in the current work, it is also necessary to inquire into the question of what kind of quantum oracle is fit for our learning-simulator in figure~\ref{fig:qlemda} in a practical manner.

Dealing with the quantum oracle is two folds: defining appropriate query function $Q$ and encoding its output $q$ on the oracle operation. The query function $Q$ maps an available inputs $x_i$ of the problem to a certain accessible values $q_{x_i}$, $Q:x_i \mapsto q_{x_i}$ ($i=1,2,\ldots,K$). Here we clarify that $Q$ is evaluated {\em classically}, and independent with the construction of the oracle operation. The finite input set $\{x_i\}$ ($i=1,2,\ldots,K$) and the query function $Q$ are determined preliminary to learning, as mentioned in section~\ref{sec:2}. 

Let us now consider a general process for oracle operation, such that
\begin{eqnarray}
\ket{j}\ket{x_i} \rightarrow e^{i \pi \varphi_{x_i}} \ket{j \oplus g_{x_i}}\ket{x_i},
\end{eqnarray}
where $\ket{j}$ is a computational basis, and $\ket{x_i}$ is a quantum state of an input $x_i$. Here, $\varphi_{x_i}$ and $g_{x_i}$ are controllable parameters depending on $x_i$. We then determine a specific form of oracle operation by choosing either $(\varphi_{x_i} = 0) \land (g_{x_i} = q_{x_i})$ or $(\varphi_{x_i} = q_{x_i}) \land (g_{x_i} = 0)$. These two types of oracle are equivalent, in the sense that they are independent with the query function $Q$, and can be converted to each other without any altering the complexity of the found algorithm \cite{Kashefi02}. In this work, we considered the latter type of oracle operation, as it is more economical in the sense that the query function is encoded into the phase without any additional system.

\section{The variants of the original $1$-bit Deutsch-Jozsa algorithm}\label{appendix_2}

In this appendix, we discuss about the original Deutsch-Jozsa algorithm and its variants for the simple case $n=1$ \cite{Bang14-2}. In such case, the learning part of our simulator consists of two single-qubit unitary operations $\hat{U}_k$ ($k=1,3$) and one oracle operation $\hat{U}_x$, as in equation~(\ref{eq:ux}). Here it is convenient to rewrite any single-qubit unitary operation $\hat{U}_k$ as
\begin{eqnarray}
\hat{U}_k(\mathbf{p}) = \exp\left(-i \mathbf{p}_k^T \boldsymbol{\sigma}\right) = \cos{\Theta_k}\hat{\identity} - i \sin{\Theta_k}\left( \mathbf{n}_k^T \boldsymbol{\sigma} \right),
\end{eqnarray}
where $\mathbf{p}_k=(p_{k,x}, p_{k,y}, p_{k,z})^T$ is a three-dimensional real vector, and $\boldsymbol{\sigma}=(\hat{\sigma}_x, \hat{\sigma}_y, \hat{\sigma}_z)^T$ is nothing but the vector of Pauli operators. Here, $\Theta_k$ is given as Euclidean vector norm of $\mathbf{p}_k$, i.e. $\Theta_k = \| \mathbf{p}_k \| = (\mathbf{p}_k^T \mathbf{p}_k)^\frac{1}{2}$, and $\mathbf{n}_k=\frac{\mathbf{p}_k}{\| \mathbf{p}_k \|}$ is normalized vector. All pure states are characterized as a point on the surface of unit sphere, called ``Bloch sphere'', and $\hat{U}_k$ rotates a pure state (i.e. a point on the Bloch sphere) by the angle $2\Theta_k$ around the axis $\mathbf{n}_k$. Such a geometric description is convenient to describe the unitary processes. 

We now turn to $1$-bit DJ algorithm $\hat{U}_1$-$\hat{U}_x$-$\hat{U}_3$ which consists of three operation steps: Firstly, the unitary $\hat{U}_1$ rotates the initial state $\ket{0}$ to a state on {\em the equator of the Bloch sphere}, i.e., $\frac{1}{\sqrt{2}}(\ket{0}+e^{i\phi}\ket{1})$, where $\phi$ is an arbitrary phase factor. The oracle $\hat{U}_x$ then flips the state to the antipodal side if $x_i$ is balanced, and leaves unchanged if $x_i$ is constant. The last unitary $\hat{U}_3$ transforms the incoming state to the corresponding output,
\begin{eqnarray}
\label{eq:out_n1}
\ket{\Psi_\text{out}(x_i)}=
\left\{
\begin{array}{ll}
\pm \ket{0}, & \text{if $x_i$ is constant}, \\
\pm \ket{1}, & \text{if $x_i$ is balanced}.
\end{array}
\right.
\end{eqnarray}
Noting that Hadamard operation $\hat{H}$ is $\pi$-rotation about the axis $\mathbf{n}=({1}/{\sqrt{2}}, 0, {1}/{\sqrt{2}})^T$, it is easily checked that the phase $\phi$ is given to be zero in the original DJ. Based on such description, we can infer that there are numerous set $\{ (\Theta_k$, $\mathbf{n}_k) \}$ ($k=1,3$) leading the initial state $\ket{0}$ to the desired output $\ket{\Psi_\text{out}(x_i)}$ as equation.~(\ref{eq:out_n1}). Thus, many variants of the original DJ algorithm exists. As an example, we give $\hat{U}_1$ and $\hat{U}_3$ found in our simulator as below:
\begin{eqnarray}
\hat{U}_1 &\simeq& \small{\left( \begin{array}{cc} {0.348 + 0.612i} & {0.631 - 0.325i} \\ {-0.631 - 0.325i} & {0.348 - 0.612i} \end{array}\right),} \nonumber \\
\hat{U}_3 &\simeq& \small{\left( \begin{array}{cc} {-0.360 - 0.609i} & {-0.031 + 0.706i} \\  {0.031 + 0.706i} & {-0.360 + 0.609i} \end{array}\right),}
\end{eqnarray}
with
\begin{eqnarray}
\small{
\left\{
\begin{array}{ll}
{\Theta_1 \simeq 0.552 \pi}, & \mathbf{n}_1 \simeq \small{({-0.243}, {0.847}, {-0.472})^T}, \\
{\Theta_3 \simeq 0.476 \pi}, & \mathbf{n}_3 \simeq \small{({0.043}, {-0.531}, {0.846})^T}.
\end{array}
\right.
}
\end{eqnarray}
The algorithm constructed with the above $\hat{U}_{1}$ and $\hat{U}_{2}$ runs as
\begin{eqnarray}
\fl
\small{
\left.
\begin{array}{ll}
\ket{0} \overset{\hat{U}_1}{\longrightarrow} {\left(\begin{array}{c} {0.704} \\ {-0.710 e^{0.18 \pi}} \end{array}\right)} \overset{\hat{U}_x}{\longrightarrow} {\left(\begin{array}{c} {0.704} \\ {-0.710 e^{0.18 \pi}} \end{array}\right)} \overset{\hat{U}_3}{\longrightarrow} \ket{\psi_\text{out}}\simeq\ket{0}, & \text{if}~x_i \in C, \\
\ket{0} \overset{\hat{U}_1}{\longrightarrow} {\left(\begin{array}{c} {0.704} \\ {-0.710 e^{0.18 \pi}} \end{array}\right)} \overset{\hat{U}_x}{\longrightarrow} {\left(\begin{array}{c} {0.704} \\ {0.710 e^{0.18 \pi}} \end{array}\right)} \overset{\hat{U}_3}{\longrightarrow} \ket{\psi_\text{out}}\simeq\ket{1}, & \text{if}~x_i \in B.
\end{array}
\right.
}
\end{eqnarray}
This algorithm is not exactly equal to, but equivalent to, the original $1$-bit DJ algorithm.


\section*{References}


\end{document}